# Unmatched Perturbation Accommodation for an Aerospace Launch Vehicle Autopilot Using Dynamic Sliding Manifolds

M. R. Saniee

*Department of Aerospace Engineering, Indian Institute of Science, Bangalore, India*

**Abstract**- Sliding mode control of a launch vehicle during its atmospheric flight phase is studied in the presence of unmatched disturbances. Linear time-varying dynamics of the aerospace vehicle is converted into a systematic formula and then dynamic sliding manifold as an advanced method is used in order to overcome the limited capability of conventional sliding manifolds in minimizing the undesired effects of unmatched perturbations on the control system. At the end, simulation results are evaluated and the performance of two approaches are compared in terms of stability and robustness of the autopilot.

## I. INTRODUCTION

To reach the aim of a mission, an aerospace launch vehicle (ALV) must move along the specified trajectory and has a required attitude. These tasks are fulfilled by the vehicle flight control system or autopilot. It forms control actions such as forces and torques to the ALV during its powered path providing the best fulfillment of the specified requirements to the vehicle terminal state vector [1].

Since dynamic equations of an ALV system can be mathematically modeled using estimated and time-varying coefficients, the most critical problem arises due to the variable characteristics of such vehicles [2]. So the attitude control systems are confronting dynamics with uncertain parameters in addition to nonlinearities and disturbances. In order to achieve an acceptable performance, robust controllers are proposed to follow the nominal trajectory [3].

One of the nonlinear robust control techniques is variable structure controls (VSC). It utilizes beneficial characteristics of various control configurations and delivers robust performance and new features that none of those structures possess on their own. The central feature of VSC is the so-called sliding mode control (SMC) on the switching manifold which the system remain insensitive to plant parameter variations and external disturbances [4].

Sliding mode control first introduces in the framework of VSC and soon became the principle operational mode for this class of control systems. Due to practical advantages such as order reduction, low sensitivity to turbulences and plant parameter variations, SMC has been known a very efficient technique to control complicated dynamic plants functioning under variable situations which are common for many processes of modern technologies. The main shortcoming of SMC namely chattering phenomenon arises due to switching in the both sides of sliding manifold. This dilemma can be treated by continuous approximation of discontinuous control or by continuous SMC design [6]. SMC also can not accommodate for unmatched disturbances unlike its powerful application for matched disturbance rejection. To obliterate this problem encountered in practice for flight dynamics and timescale separation of central loops in multi-loop systems, dynamic sliding mode (DSM) has received considerable attention [7]. DSM exploits benefits of dynamic compensators to offer a switching manifold in order to provide the systems with robustness characteristics against unmatched perturbations [8]. This technique can be applied in variety of complex control systems and even automated passive acoustic monitoring devices used for studying marine mammal vocalizations and their behaviors [9-12]. In this analysis enhanced properties of a control, designed based on DSM for longitudinal channel output tracking of a time varying ALV will be demonstrated in comparison to that of CSM control.

Section 2 and 3 presents CSM and DSM control theory, respectively. ALV dynamics are offered in Sec. 4. CSM and DSM autopilot designed and simulation results demonstrated in Sec. 5. Section 6 devoted for conclusion.

## II. CSM CONTROL THEORY

Consider the following dynamic model:

$$\dot{x} = \mathbf{A}x + \mathbf{B}u + \mathbf{F}(\mathbf{x},t) \qquad (1)$$

where $\mathbf{x}(t) \in R^n$ and $u(t)$ are the state vector are the control vector, respectively; A and B are constant matrices; $F(\mathbf{x},t)$ is a bounded disturbance; It is assumed that {A,B} is a

controllable pair and rank(B)=$m$. The conventional sliding surface $S(x,t)$ can be defined as:

$$S(x,t) = (\frac{d}{dt} + \lambda)^{n-1} e \quad (2)$$

where $\lambda$ is a positive real constant and $e$ is tracking error as:

$$e = x - x_d \quad (3)$$

where $x_d$ is the nominal trajectory and $e(0)=0$. In this study, it is assumed that $n=2$ and sliding surface can be determined in terms of error as follows:

$$S = \dot{e} + \lambda e \quad (4)$$

The tracking error will asymptotically reach zero with a control law of bellow form:

$$u = \begin{cases} u^+, & S(x,t) > 0 \\ u^-, & S(x,t) < 0 \end{cases} \quad (5)$$

So, the system dynamics moves from any initial states toward the sliding hyperplanes in a certain amount of time and maintains on it hereafter [4]. In other words, the existence of the conventional sliding mode in the designed sliding manifold is provided.

This two-stage design becomes simpler for systems in so-called regular form. Since rank{B}=$m$, matrix B in Eq. (1) can be partitioned as:

$$\mathbf{B} = \begin{bmatrix} \mathbf{B}_1 \\ \mathbf{B}_2 \end{bmatrix} \quad (6)$$

where $\mathbf{B}_1 \in \mathbf{R}^{(n-m) \times m}$ and $\mathbf{B}_2 \in \mathbf{R}^{m \times m}$ with $\det(\mathbf{B}_2) \neq 0$. The nonsingular coordinate transformation,

$$\begin{bmatrix} x_1 \\ x_2 \end{bmatrix} = \mathbf{T}x \quad , \mathbf{T} \in \mathbf{R}^{n \times n} \quad (7)$$

converts the system Equation (1) to regular form:

$$\begin{cases} \dot{x}_1 = \mathbf{A}_{11} x_1 + \mathbf{A}_{12} x_2 + \mathbf{F}_1 \\ \dot{x}_2 = \mathbf{A}_{21} x_1 + \mathbf{A}_{22} x_2 + u + \mathbf{F}_2 \end{cases} \quad (8)$$

where $x_1 \in \mathbf{R}^{n-m}$, $x_2 \in \mathbf{R}^m$, $\mathbf{A}_{ij}$ are constant matrices for $i, j = 1, 2$, $F_1$ is unmatched disturbance and $F_2$ is matched disturbance.

Lyapunov direct method is used to obtain the control law. A candidate function is selected as:

$$V = \frac{1}{2} s^T s \quad (9)$$

with $V(0) = 0$ and $V(s) > 0$. The condition, guaranteeing an ideal sliding motion, is the $\eta$-reachability condition given by:

$$\frac{1}{2} \frac{d}{dt} s^2 \leq -\eta |s| \quad (10)$$

where $\eta$ is a small positive constant. Therefore, a control input can be chosen as:

$$u = (\lambda \mathbf{B})^{-1} (-\dot{e} - \lambda \mathbf{A} x + \ddot{x}_d - \rho sign(s)) \quad (11)$$
$$\equiv u_{eq} + u_{disc}$$

where $\rho$ is positive real constant, $u_{eq}$ is the continuous control which is called "equivalent control" and $u_{disc}$ is the discontinuous control that switches around the sliding surface and so, system state synchronously moves on this manifold and toward the origin.

## III. DSM CONTROL APPROACH

The main characteristic of the dynamic sliding mode is being compensator. It means that DSM control designs control law of each step based on previous step data and so, the system may need some additional dynamics to improve the system and sliding mode stability besides the desired system response.

Dynamic sliding manifold can be modeled as a linear function in terms of some states and tracking error as follows [7]:

$$\Im(x_2, e) = x_2 + W(s)e \quad (12)$$

where $x_2 \in R^m$, $s = d/dt$, $W(s) = \frac{P(s)}{Q(s)}$ and $P(s), Q(s)$ are polynomials of $s$. The operator $W(s)$ has to be specified in order to provide the desired plant behavior as well as rejecting effects of unmatched disturbance $\mathbf{F}_1$.

To ensure the occurance of the sliding mode on the sliding manifold (12), the discontinuous control should be designed. By using the Lyapunov's function (9) and the reachibility condition (10), the control law can be given as:

$$u = -A_{21} x_1 - A_{22} x_2 - W(s)\dot{e} - \rho sign(\Im) \quad (13)$$
$$\equiv u_{eq} + u_{disc}$$

The existence of the sliding mode in the dynamic sliding surface can be proven if $\Im(x_2, e) = 0$, derivative of (9) can be identified as:

$$\dot{V} = \Im^T \dot{\Im} = \dot{x}_2 + W(s)\dot{e} \quad (14)$$

Substituting (8) and (13) into (14) should yield the following expression:

$$\dot{V} = -\rho sign(\Im) < 0 \quad \forall \rho > 0 \quad (15)$$

Consequently, the surface $\Im = 0$ is attractive and DSM control provides asymptotic stability to the states of tracking error dynamics.

## IV. THE EQUATIONS OF MOTION FOR ALV

Newton's second law can be employed to extract the motion equations of an ALV. Assuming rigid airframe for the vehicle, the 6DOF equations of motion obtained as follows [2]:

$$F_x = m(\dot{U} + qW - rV)$$
$$F_y = m(\dot{V} + rU - pW)$$
$$F_z = m(\dot{W} + pV - qU)$$
$$M_x = I_x \dot{p} \quad (14)$$
$$M_y = I_y \dot{q} + (I_x - I_y)pr$$
$$M_z = I_z \dot{r} + (I_y - I_x)pq$$

Since altitude control systems of an ALV are usually simplified into a linear set of dynamical equations, a linear ALV model is developed in this article. Considering small perturbations, linearized equations of motion can be obtained as follows [1]:

$$\dot{v}_z = Z_v v_z + Z_q q + Z_\theta \theta + Z_{\delta e} \delta_e$$
$$\dot{q} = M_{vz} v_z + M_q q + M_{\delta e} \delta_e$$
$$\dot{v}_y = Z_v v_y + Z_r r + Z_\theta \theta + Z_{\delta r} \delta_r \quad (15)$$
$$\dot{r} = M_{vy} v_y + M_r r + M_{\delta r} \delta_r$$
$$\dot{p} = M_p p + M_{\delta a} \delta_a$$

where $Z, M$ are dynamic coefficients and $\delta$ is deflection of trust vector.

Since the control objective is to track guidance command in pitch channel, thus the two first equations will be regarded as required dynamics and the other three ones are waved belonging to yaw and roll channels. Time varying coefficients of pitch dynamics in Eq. (15) are shown as in Figure 1.

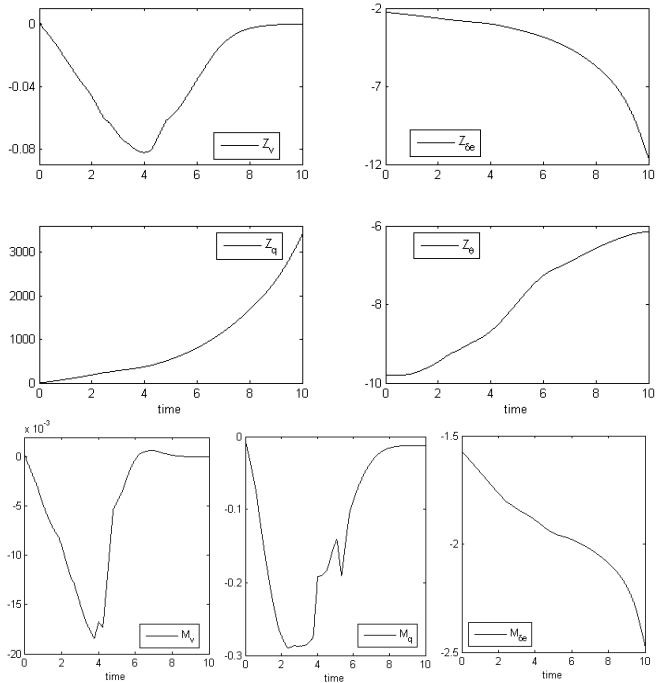

Figure 1. Longitudinal dynamic coefficient

The trust vector deflection of servo dynamics can be described as:

$$[TF]_{servo} = \frac{\delta}{\delta_c} = \frac{1}{0.1s + 1} \quad (16)$$

with a rate limit of $|\frac{d\delta}{dt}| < 25 \deg/\sec$. Because reference signal is pitch rate, a rate gyro is selected as follows:

$$[TF]_{gyro} = \frac{(80\pi)^2}{s^2 + (40\pi)s + (80\pi)^2} \quad (17)$$

## V. AUTOPILOT DESIGN AND SIMULATION RESULTS

In this section, both CSM and DSM control designed for the time varying ALV pitch longitudinal channel and excellent performance of DSM in comparison to that of CSM are demonstrated.

### A. CSM Control Design

The goal is to generate the control $\delta_e$ to enforce state motion on CSM:

$$S = \dot{\theta}_e + K\theta_e \quad (18)$$

where $\theta_e = \theta_c - \theta$ and $K = const.$ is chosen in order to make ALV track the commanded pitch rate $q_c$ and so, the states trajectory of system asymptotically converge to the sliding manifold $S = 0$.
Using Lyapunov function (9), its derivative is derived as:

$$\dot{V} = S^T \dot{S} = S^T[\dot{q}_c - M_{vz} v_z - M_q q - M_{\delta e} \delta_e + K\dot{\theta}_e] \quad (19)$$

By utilizing the equality form of (10) for ensuring asymptotic stability of the system, the necessary control is given as:

$$\delta_e = M_{\delta e}^{-1}[\dot{q}_c + Kq_e - M_{vz} v_z - M_q q + \rho sign(S)] \quad (20)$$

where $K = 1$ and $\rho = 0.01$ have been selected.

The control law (20) is discontinuous and will cause chattering on the manifold (18). To solve this undesired phenomenon, the discontinuous term $sign(S)$ in Eq. (20) is replaced by the continuous term $sat(S/\varepsilon)$, where $\varepsilon$ is a real small constant whose value is chosen $10^3$ in this research.

### B. DSM Control Design

Following the procedure in [7], the design procedure for dynamic sliding manifold is presented. Note that to transform the ALV longitudinal equations of motion to regular form (8) and avoiding singularity, the servo dynamic equation (16) is added to plant equations. Thus, $\delta$ is converted to one of system states and $\delta_c$ will be the control effort.

Based on Eq. (12), the following expression for dynamic sliding manifold is defined:

$$\Im = \delta + W(s)e \quad (21)$$

which $W(s)$ can be selected as bellow:

$$W(s) = \frac{a_1 s^2 + a_2 s + a_3}{b_1 s^2 + b_2 s} \quad (22)$$

where $a_1, a_2, a_3, b_1, b_2$ are real indices determined for each iteration. In order to obtain these coefficients, tracking error achieved as:

$$e = \frac{q_c}{1 - W(s)G(s)} \quad (23)$$

where $G(s)$ is the transfer function of $q(s)$ relative to $\delta(s)$. By comparing characteristics equation for (23) and integral of time multiplied by absolute tracking error criterion with:

$$s^5 + 2.8 w_n s^4 + 5 w_n^2 s^3 + 5.5 w_n^3 s^2 + 3.4 w_n^4 s + w_n^5 = 0 \quad (24)$$

where $w_n = 10 Hz$ is chosen and related parameters identified at each moment. Also, the control $\delta_c$ is given as follows:

$$\delta_c = -\rho sat(\Im/\varepsilon) \quad (25)$$

where $\rho = 1$ and $\varepsilon = 10^3$ is considered.

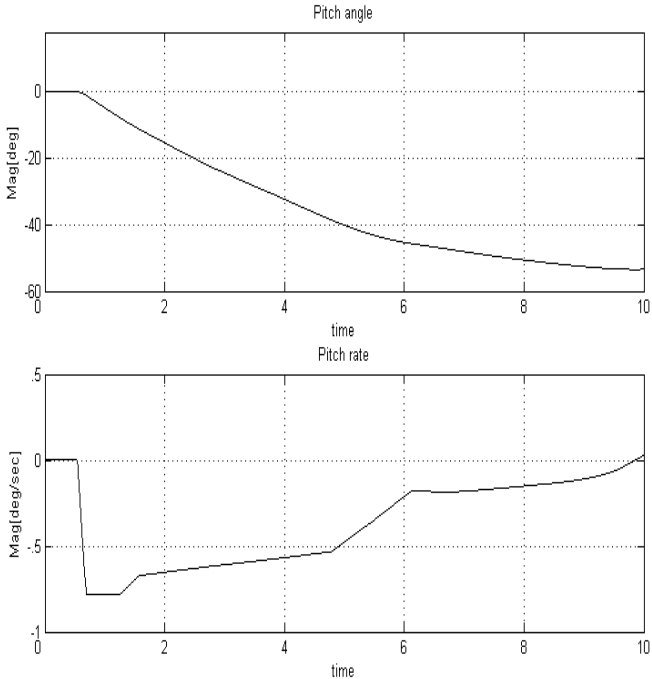

Figure 2. Desired pitch and pitch rate to be tracked

In this paper, pitch rate program has been designed offline as shown in Figure 2 and it is desired to be tracked during the flight envelope. Since sliding mode control can properly accommodate for matched disturbances [4], simulation was run in presence of unmatched disturbances depicted in Figure 3 such that $f_{11}$ and $f_{12}$ are exerted to the first and second expressions in Eq. (15), respectively.

The simulation results with dynamic and conventional SMC are illustrated in Figure 4 and Figure 5, respectively. It is shown that DSM unlike the CSM can follow the nominal trajectory very closely and withstands the unmatched perturbations.

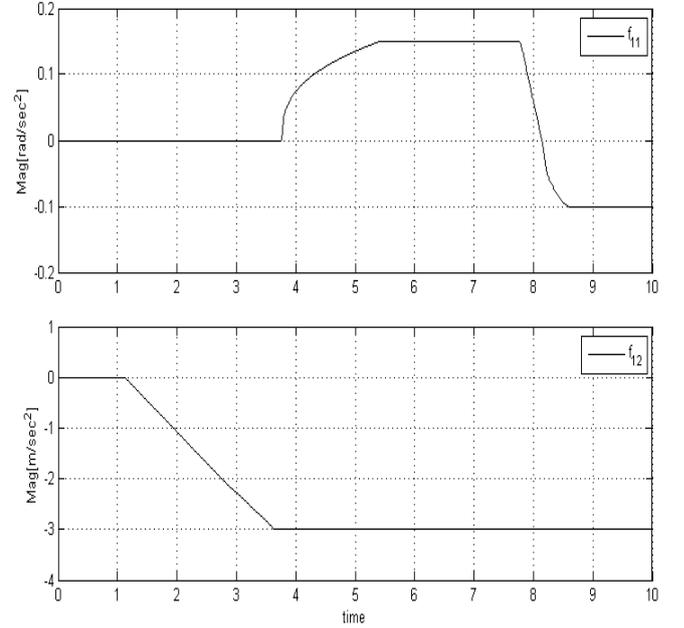

Figure 3. Unmatched disturbances profiles

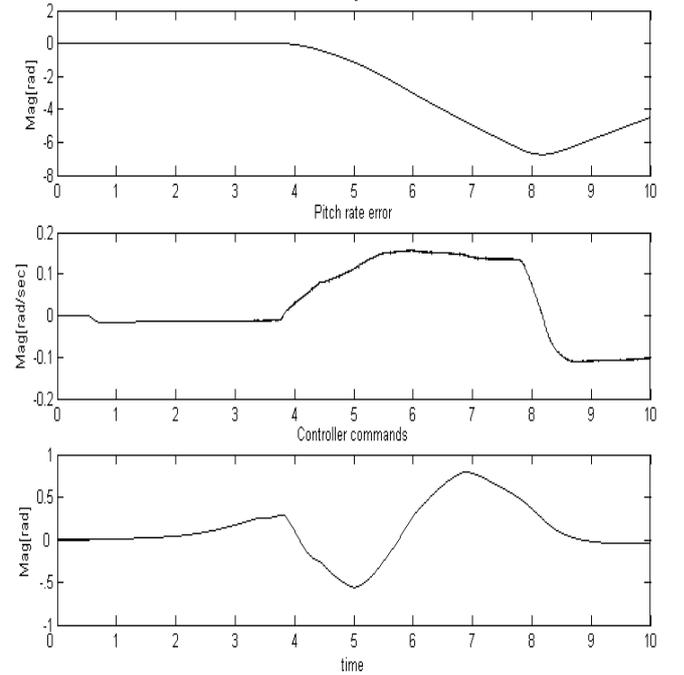

Figure 4. Pitch angle error, pitch rate error and controller command obtained from CSM autopilot

In this research, it is assumed that $a_1 = b_1 = 1$ and the other three coefficients are determined by the corresponding algorithm at each moment during the ALV flight time whose variations are illustrated in Figure 6.

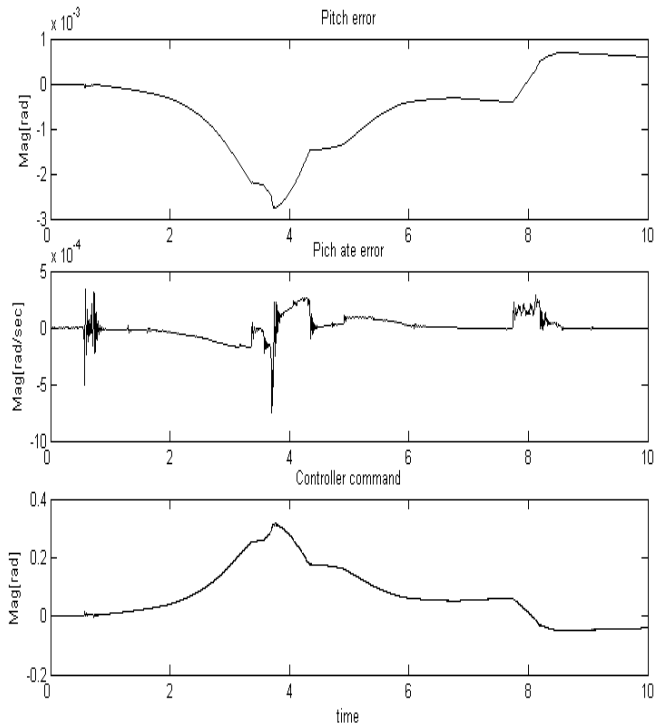

Figure 5. Pitch angle error, pitch rate error and controller command obtained from DSM autopilot

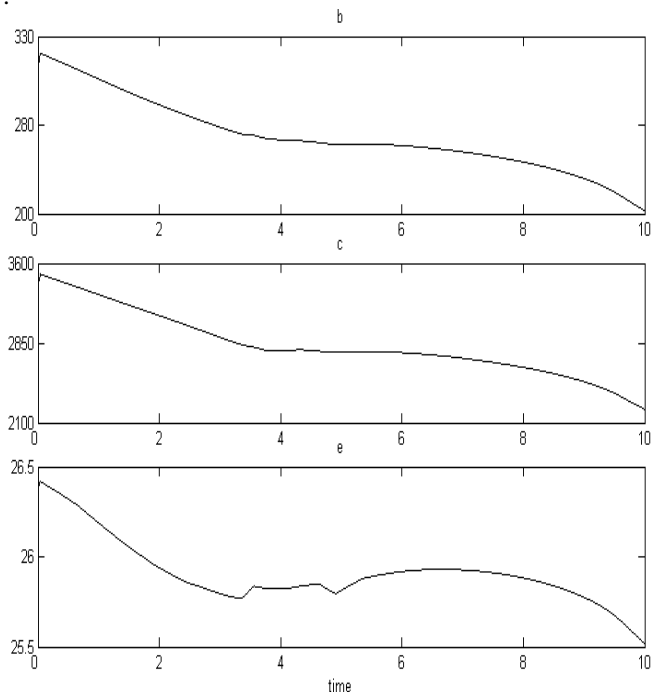

Figure 6. Indices variations of $W(s)$ calculated on-line

## VI. CONCLUSION

In this research, commanded pitch rate tracking with unmatched disturbances for the atmospheric flight of a time-varying ALV is considered in SMC. Both conventional and dynamic SMC was designed and closed-loop system operations of these methods were compared.

Results show that dynamic SMC can accommodates unmatched disturbances and output tracking errors is much less than those of CSM, while conventional SMC does not operate properly and cannot satisfy requirements of system performance. The simple and straightforward design procedure, together with the encouraging robustness against unmatched disturbances; invite further application of this approach.